# IoT Forensic Frameworks (DFIF, IoTDOTS, FSAIoT): A Comprehensive Study

**Mohammad A. Hassan, Ghassan Samara, and Mohammad Abu Fadda**

Computer Science Department, Zarqa University, Zarqa- Jordan
E-mail: mohdzita@zu.edu.jo
Computer Science Department, Zarqa University, Zarqa- Jordan
E-mail: gsamara@zu.edu.jo
Computer Science Department, Zarqa University, Zarqa- Jordan
E-mail: Mohm.fadda@gmail.com

**Abstract**

***In the Internet of Things, millions of electronic items, including automobiles, smoke alarms, watches, eyeglasses, webcams, and other devices, are now connected to the Internet (IoT). Aside from the luxury and comfort that the individual obtains in the field of IoT, as well as its ability to communicate and obtain information easily and quickly, the other concerning aspect is the achievement of privacy and security in this connection, especially given the rapid increase in the number of existing and new IoT devices. Concerns, threats, and assaults related to IoT security have been regarded as a potential and problematic area of research. This necessitates the quick development or creation of suitable technologies with the nature of crimes in the IoT environment. On the other hand, criminal investigation specialists encounter difficulties and hurdles due to various locations, data types, instruments used, and device recognition. This paper provides an in-depth explanation of the criminal content of the Internet of Things. It compares its stages to the detailed stages of traditional digital forensics in terms of similarities and differences, the frameworks used in dealing with electronic crimes, and the techniques used in both types. This paper presents previous discussions of researchers in the field of digital forensics. For the IoT, which brings us to the most important parts of this paper, which is a comprehensive study of the IoT criminal frameworks that are used to protect communication in the field of IoT, such as Digital Forensic Investigation Framework (DFIF), Digital Forensic Framework for Smart Environments (IoTDOTS), Forensic State Acquisition from the Internet of Things (FSAIoT), and discusses the challenges in their general frameworks and provides solutions and strategies.***

# 1 Introduction

As a result of technological breakthroughs and the internet, the amount of information and data available today is increasing. Furthermore, the Internet has impacted the world that it has become one of the highest priorities for facilitating life affairs. The traditional use of the Internet has expanded via two dimensions: time and space, allowing for wireless connectivity to mobile devices worldwide at any time. However, because of the amazing development of the present digital world, any object may be converted into a digital model and has the power to connect wirelessly. This digital technology is composed of diverse things connected to the so-called IoT. The IoT's aspiration reaches the construction of sophisticated, intelligent, self-connected systems without human intervention to achieve common goals [1, 2, 3, and 4].

IoT is a big draw for many scientific researchers and diverse technical company sectors, and this draw is due to the tremendous capabilities of these approaches. Many services and applications are shared between varied devices in wired and wireless connectivity via IoT gadgets [5]. It can also cope with massive amounts of data in order to improve the efficiency of communication between heterogeneous organs and the ability to outperform human operations [6, 7, and 8]. The goal is to configure systems employing distant sensors and diverse devices to identify and profit from such vast amounts of data.

In recent years, there has been a tremendous increase in the adoption of IoT technologies. At the same time, these IoT-based smart gadgets have been used in significant industries such as healthcare, transportation, cellphones, smart cities, etc. The goal of IoT is to make people's lives more adaptive and dynamic. The Internet of Things (IoT) business, for example, is anticipated to increase from $892 billion in 2018 to $4 trillion by 2025 [9]. Machine-to-Machine (M2M) connection is employed in various applications, including smart cities and transportation [10, 11].

The advent of the Internet of Things aided in advancing technology to unprecedented levels. The components of the IoT are what we use from many smart gadgets in many facets of our lives. However, with every great concept comes a set of risks. IoT machines exchange data with millions of different devices all over the world [12]. With such a large-scale link, they are especially looking for persons with criminal motives and malicious assaults. Because the potential hazards of cyber security are so great, it is vital to be prepared with the tools needed to tackle this risk and avoid it as much as possible. As a result, forensics is critical for IoT [13,14].

This paper aims to provide a high-level overview of IoT forensics and strategy. First, we define and discuss digital forensics and IoT forensics. The IoT forensic framework is then described (DFIF, IOTDOTS, and FSAIoT). Third, we discuss related work that uses forensic methodologies to create various applications. Fourth, we build a list of IoT forensic issues and prospects. Finally, we provide findings and recommendations for the future.

### 1.1 Defining Digital Forensics

Digital forensics is a branch of forensic science concerned with investigating cybercrimes by processing digital data and information to get evidence. The purpose of these evidences after they have gone through systematic stages of discovery, examination, collection, and storage, as well as taking care during these stages to maintain the integrity and effectiveness of evidence when presented as legally considered evidence before competent courts [15, 16].

Several writers have provided definitions of digital forensics, including Servida in [17, 18]. The authors presented digital forensics as a concept associated with digital devices through defining, evaluating, and acquiring digital evidence gathered from these devices. The issue is not limited to computers, but also includes devices such as phones, smart devices, cameras, network devices, and so on, and here is an introduction to digital forensics for the IoT environment, while other researchers and an example of that in [19, 20] gave a simple illustration of digital forensics that the presence of the ability to provide digital evidence stored on digital devices and the advantage of this evidence has strong facts by identifying

In general, digital forensic is separated into various classes that differ in the content of information and how to deal with itThese . classes are: forensic for computer crimes, forensic for network crimes, forensic for IoT, and forensic for smartphone crimes and forensic for clouds.

## 1.2 IoT Forensic

The investigation area of IoT in theoretical survivor comes from a branch of the Digital Forensic, which means investigating crimes in the real and virtual worlds, but in the IoT are more broadly and deeply because it deals with their dazzling and diverse data in quality and quantity. IoT has also contributed to a rapidly developing development that has addressed obstacles and security issues in Digital Forensic that differ from  theprevious status. This is appealing attention for proprietors of breakthroughs, sequences, and espionage [21]. It is obvious from the preceding that it is very difficult to follow traditional Digital Forensic mechanisms and procedures, as well as the need to develop advanced models and technical methods in response to the development of the world's IoT in all essential stages of investigation [22, 23], whether to contain digital evidence and analyzes, which are for cloud services and regulations in the Internet environment. As indicated in Fig 1, IoT investigations are divided into three axes: digital forensic on the device level, digital forensic on the network level, and digital forensic on the cloud [24, 25].

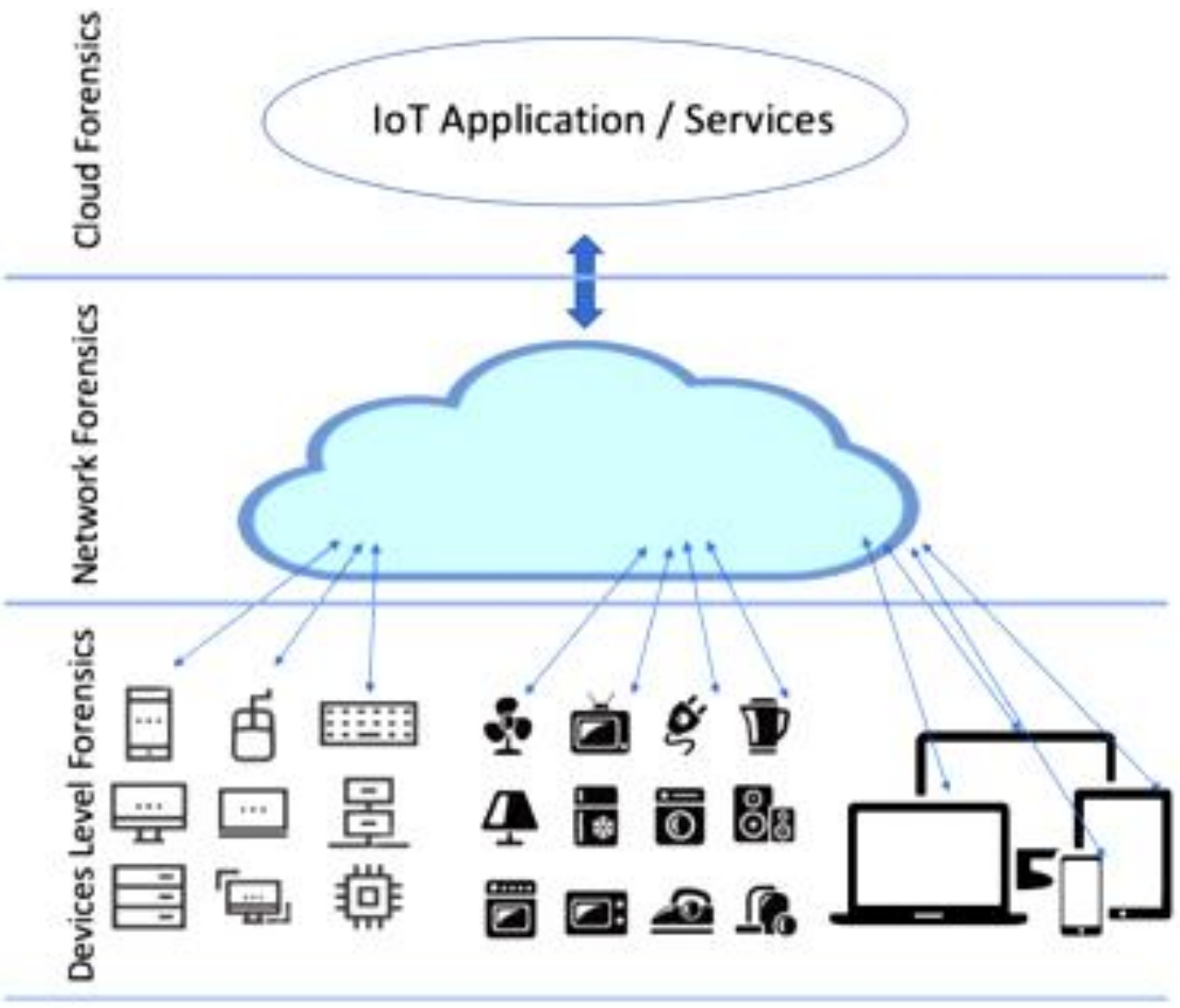

Fig. 1: IoT forensic types.

- **Device Forensics:** Physical devices are recognized as the major source of evidence for a given cybercrime at this level. The investigator identifies the targeted IoT device and collects the necessary evidence, which can take many forms such as images, video, audio, and so on. Mobile forensics is a type of device forensics that can be used to save personal data such as details, photographs, documents, and notes, as well as SMS and MMS messages [26].
- **Network Forensics:** At this level, the types and forms of communication networks that connect various devices in the IoT environment are explained, which include Personal Area Networks (PAN), Local Area Networks (LAN), and Large Area Networks (LAN) (WAN). To implement the digital investigative process in the previously described networks, recognizing the sources of various assaults and extracting information on access through network access and departure movements, for legal evidence through traceability for individuals moving within these networks [27].
- **Cloud Forensics:** The Internet cloud is one of the most important stages in the digital investigation process because data and information received from Internet devices are equipped and stored on this cloud, despite the limited storage and computation capabilities of these devices, making the cloud a very rich source of the digital investigation process. This has already improved the fact that the level of layers is the primary model in the inquiry process[28].

## 2 Literature Review

In order to adapt digital forensics to the IoT system, several recent studies have proposed new investigative models or examined emerging difficulties in IoT forensics. This section includes a comprehensive analysis of current IoT forensics research that can help digital investigators and specialists in the area maintain track of recent studies and introduce new ones.

Hegarty and colleagues addressed the complexity of digital forensic in IoT in [29], and the authors proposed a cloud computing implementation approach for digital inquiry. A general analysis of proposed solutions and a system structure are all part of the effort. However, they have not offered a strategy for effecting their notion.

Others in [30] provided a method for obtaining IoT digital proof. Their dissertation depicts a theoretical framework for IoT forensics. Various investigative measures are depicted in order to obtain data for further research.

The researchers addressed forensic topics in the paper [31] and explored the contrasts and similarities of forensic investigation departments from an ancient and modern perspective. And they came across several limits and challenges for forensic specialists in IoT smart devices. Furthermore, they offered a suggestion that emphasizes the forensic foundation for smart device technologies in terms of mobility and reinforces the grounds for the necessity to use the information retrieved from sensors and the processes that occur on them. The approach they proposed was theoretical and was not transferred and implemented to assess its efficiency practically.

In [32], the researchers discussed in detail the obstacles and challenges confronting the IoT forensic investigation processes in the era of the massive spread of this virtual world. Hence, they included in this research the difficulties in applying the foundations of digital forensics in the stages of identifying data of both physical and logical types and obtaining them in the stages of analysis. The authors of this study presented a proposed model that integrates forensic in its two forms, forensic on the cloud side and forensic on the user side, including concerning the devices involved and used, to obtain a framework that supports and addresses digital investigation processes and the issues that arise as a result of digital forensics. They presented proposals to build digital standards specified by forensics used for Internet of Things Applications (IoTA) and promote IoT-based research.

Another study completed by [33] discussed the threats and opportunities in the fields of IoT monitoring and digital forensics. They briefly reviewed general monitoring and crime scene investigation difficulties regarding protection and threats. They focused on the issues of privacy, security, and forensics in the IoT era.

Using new investigative models developed by scholars such as Oriwoh et al. [34]. These models demonstrated high effectiveness in the investigation process by dividing the attack areas into three areas, using scenarios to get evidence. These scenarios were produced based on a practical study of criminals who used a new strategy in conducting their electronic crimes. After analyzing and interpreting the data from these studies, a model was developed that uses areas as a foundation for investigating the IoT environment and revolves around three main axes: Area 1, the internal network, and Zone 2, the parties, software, and hardware used within the network's boundaries. Zone 3 includes any parties, software, hardware, and devices utilized outside the network's bounds.

Shrivastava et al. [35] have also developed a forensic analysis methodology for computer investigators to capture or retain vital evidence from a wide range of computer interfaces. Their methodology was centred on guaranteeing information security through implementing organizational policies, best practices, and training and taking the required safeguards to prevent data leakage.

Sathwara et al. [36] highlighted creating a digital model structure for evidence collecting in IoT digital forensics. The goal was to study the components and methodological methods used in the IoT area in terms of the considered evidence collection stage and get practical answers to the issues in this stage. Finally, the researchers developed an ecosystem to aid investigators at the stage of acquiring evidence and addressing it in the IoT environment.

# 3 IOT Forensic Framework

***A. Digital Forensic Investigation Framework (DFIF):***

(DFIF-IoT) is a forensic framework that can provide potential IoT investigators with a level of assurance. This framework provides the following advantages: It complies with ISO/IEC 27043: 2015, an international information technology convention, in order to efficiently prosecute computer crime in court [37]. This section describes the processes taken to create the Digital Forensic Investigation Framework (DFIF):

- Step 1. Recognize current structures
- Step 2. Formation of phase name
- Step 3. Mapping the Process

***B. Digital Forensic Framework For Smart Environments (IoTDOTS):***

IoTDots is a new digital forensic architecture for intelligent surroundings such as smart offices and smart homes. IoTDots has two basic stages, as indicated in fig.2 [38]: stage one IoTDots-Modifier and stage two IoTDots-Analyzer. IoTDots Modifier scans smart app source code at build time, detecting forensically relevant information and automatically inserting monitoring logs. During the execution phase, the logs are saved in an IoTDots database. The IoTDots-analyzer then use data analysis and machine learning techniques to extract useful and accurate forensic information from the device's usage.

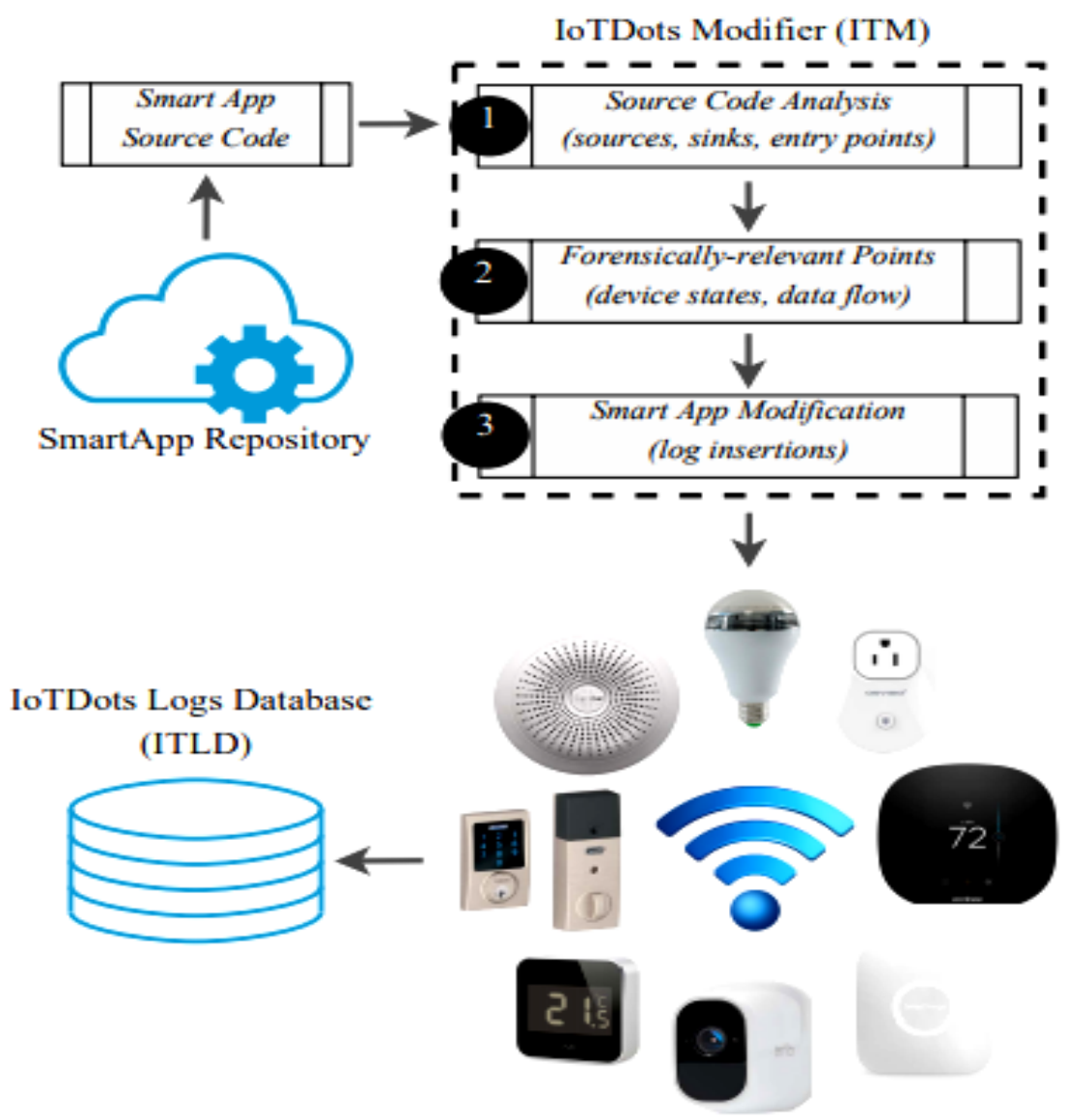

Fig. 2: IOTDOTS modifier and analyser.

To identify the perpetrator of the crime and hold him accountable, the general framework of the work must be able to identify information with a high criminal value, such as the user's natural movements and abnormal movements, as well as offensive and suspicious activity of applications and individuals [39]:

• Natural user movements: moves that occur through user accounts while adhering to the relevant security guidelines.

• Abnormal user movements: Any activity that occurs due to authorized user accounts' negligence and in breach of approved security guidelines.

• Attacking and suspicious activity: Any activity intended by allowed or unauthorized user accounts or smart devices and applications that intentionally breaches security guidelines and endanger others and the system's status.

As a result, IoTDots can be used criminally based on (1) unusual user movements and (2) offensive and suspicious conduct. The following activities are studied with respect to time [40] to see the potential of IoTDots in spotting any irregular movements of users:

**A) Time-independent activities:**

- Activity No. 1: The authorized user's ability to manage smart devices unlawfully within the scope of the smart environment at any time.

- Activity No. 2: The authorized user's ability to monitor smart devices in an unauthorized manner beyond the boundaries of the smart environment at any time.

- Activity No. 3: Within the scope of the smart environment, the potential of locating an authorized user in a banned region.

**B) Time-dependent activities:**

- Activity 4: Move throughout the smart environment outside of normal time, whether the user is dependent or not.
- Activity 5: Access to the smart environment by the user, whether permitted or not, at an inconvenient time.

Because IoTDots defines smart environment security protocols, user activities are categorized as time-dependent or non-time-dependent. Users' movements are recorded, and their performance has nothing to do with IoTDots due to the inability to manage it because it is one of the procedures followed by the first party. Activities (1, 2, 3) are time-independent since they do not use time as a basis. On the other hand, any activity with a set completion period is accurately included by using particular security measures for the security environment. And the two activities (4 and 5) are included in this group for maximum criminal advantage in a short period.

**C. Forensic State Acquisition from the Internet of Things (FSAIoT):**

FSA IoT is a holistic system that collects data in three states: IoT unit, cloud, and controller [41].

The FSA IoT is composed of two parts: the first is the Forensic State Acquisition Control Unit, which is a central control unit, and the second is a collection of procedures for collecting the case, which is expressed as the current state of the IoT device, as follows [42]:

1) ***Controller to IoT device***: As illustrated in Fig 3, most devices in the IoT ecosystem are depicted as a simple control unit coupled to device control. An example of this is an IP camera controlled by a network device. When any movement is detected, the difference in the situation is identified and reported to the listening control unit. Other steps are then conducted depending on the monitoring of this change.

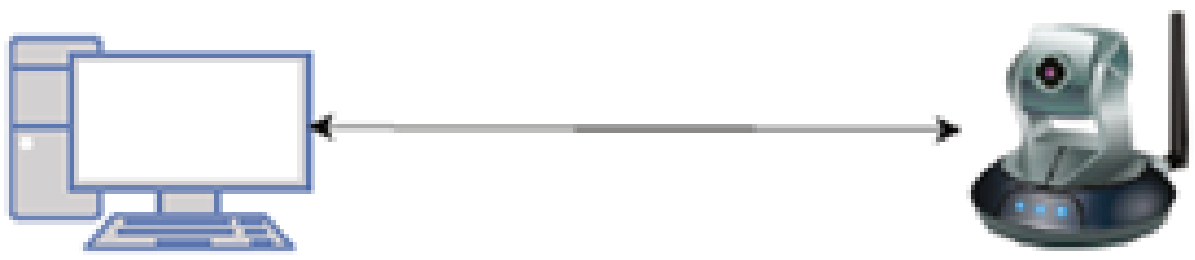

Fig. 3: controller to device.

2) ***Controller to cloud*:** Now, many internet gadgets use the cloud's services as controllers and to receive data. In this situation, the application software interface responsible for administering the IoT device can retrieve the current state of any IoT device via cloud data. The API8 of the Nest is a wonderful illustration of this. Communications to the nest temperature controller are accessed via cloud calls, with this access fig.4, the possibility of being able to access the nest device or designing individual programmed parts to obtain cloud data directly to monitor any changes in the state and state of the IoT device, and the cloud is responsible for controlling it.

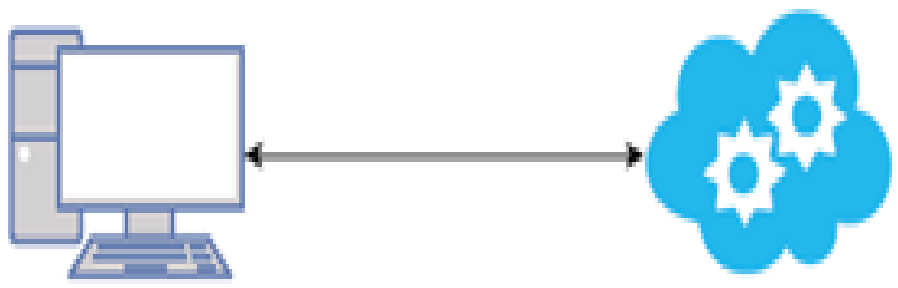

Fig. 4: controller to device.

*3)* ***Controller to controller***: *The units in charge of controlling the Internet are linked to easy control. This link is made possible via mobile applications or online interfaces, as a consequence of which various system statuses can be obtained. From this point in charge of data collecting, it became evident how beneficial our system would be. When businesses include IoT devices into their content, the necessity for centralized control points grows.*

# 4 Iot Forensic Challenges and Proposed Solutions

While the Internet of Things provides investigators with a rich environment with a wide range of characteristics, some factors can stymie an effective investigation in the IoT environment. The following are some of the challenges, as well as possible solutions for each of them:

- **Data Format:** The data format delivered by IoT devices differs from that stored in the cloud. Furthermore, users do not have direct access to their data, and the data appears in a different format from that in which it is kept. Before being processed on the Cloud, data can be analyzed using analytic tools. As a result, before doing analysis, the data type must be returned to its original format in order to be admissible in a legal proceeding [43].
  **Suggested Solution:** This issue can be remedied by creating a modern standard data format for IoT suitable for use in a court of law. Before carrying out the evaluation, the data should be returned to its original format.

- **Data Location:** Due to the restricted storage capacity in devices used for communication and electronic connectivity in all its forms in the IoT, this complicates and greatly complicates the process of data monitoring, and the global expansion of private IoT networks adds to the difficulty. An example of this is an IoT device located in one location whose data is uploaded to a cloud in another region. As a result, various requirements are taken into account [44].
  **Suggested Solution:** Preparing dynamic servers specifically designed to monitor the location of IoT devices so that if the structure used in the IoT changes, it is moved to another server that can contribute to reaching the location of the device used in the crime and thus obtaining information that serves the investigation process more effectively.

- **Forensic Tools:** Because of the variances in uniformity across the devices used and dispersed in IoT networks, this proves the inadequacy of traditional methodologies employed in digital forensics. The vast volume of digital evidence collected from IoT devices causes challenges of many types throughout the evidence-gathering stage and affects the capacity to distinguish between hacked devices. In a normal situation, the evidence presented in court must be considered. The need for it verified, and the low degree of safety present in the IoT does not serve this goal, as the foregoing strongly requires the presence of advanced technologies and tools that can keep pace with the nature of the IoT environment. The methods save time for investigators and the effort to reach fruitful investigation results in the world of the IoT [45].
  **Suggested Solution:** Investigating Internet crimes necessitates the participation of specialists with a vast understanding of IoT to help judge these crimes in court. The problem of tools utilized can be handled by working with these specialists to design tools that the specialists have already approved in the courts that handle these matters.

- **Device Identity:** The inquiry of the crime is deliberate regarding the existence of a criminally committed crime. Cloud services are based on the lack of identification that reveals its owner in user accounts [46, 47], which does not assist in acquiring digital evidence that allows access to the perpetrator's identity of the crime. In other words, in the context of investigating IoT crimes, retrieving digital evidence from the cloud does not imply that this data can be utilized to track either the culprit or the device used in the crime.
  **Suggested Solution:** Reconsidering the approval of user accounts, whether when the device is manufactured or when it is connected to the cloud, in order to impose the use of useful and brief identifying information for users on the cloud, contributing to a proposed solution to the issue of the user device's identity and the information associated with it.

# 5 Conclusion

The document initially established the terms Internet of Things (IoT), Digital Forensic, and IoT Forensic, among others. As a bonus, various new approaches, such as DFIF, DOTS, and FSA, have been investigated inside this framework. As a matter of fact, the vast majority of them are mostly focused on adapting traditional forensics techniques to IoT forensics processes. The study concludes by discussing many issues that the IoT forensic framework faces, including those related to the data format, forensic tools (including those for identifying data), data location, and data identity. We also provided solutions to these challenges.

## Notes on contributors

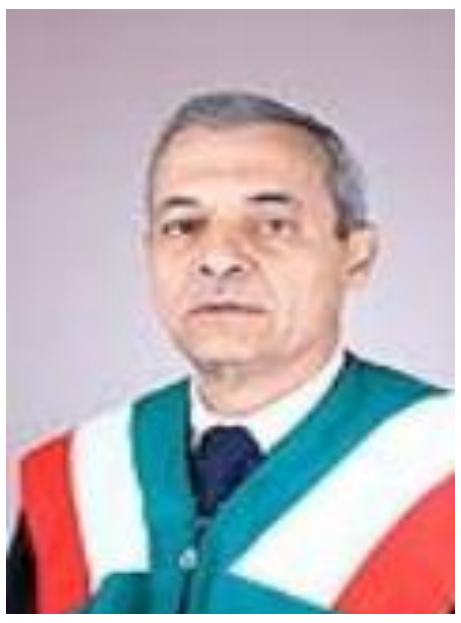

***Mohammad Hassan*** has received his BS degree from Yarmouk University in Jordan in 1987, the MS degree from University of Jordan, in 1996, and the PhD degree in computer information systems from Bradford University, UK in 2003. He is working as an associate professor in the department of computer science at Zarqa University in Jordan. His research interest includes information retrieval systems and database systems.

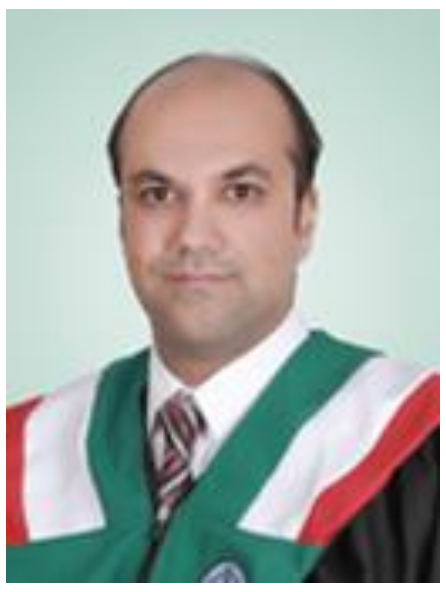

***Ghassan Samara*** *Holds BSc. and MSc. in Computer Science, and PhD in Computer Networks. He obtained his PhD, from Universiti Sains Malaysia (USM) in 2012. His field of specialization is Cryptography, Authentication, Computer Networks, Computer Data and Network Security, Wireless Networks, Vehicular Networks, Inter-vehicle Networks, Car to Car Communication, Certificates, Certificate Revocation, QoS, Emergency Safety Systems. Currently, Dr. Samara is an associate professor at Zarqa University, Jordan.*

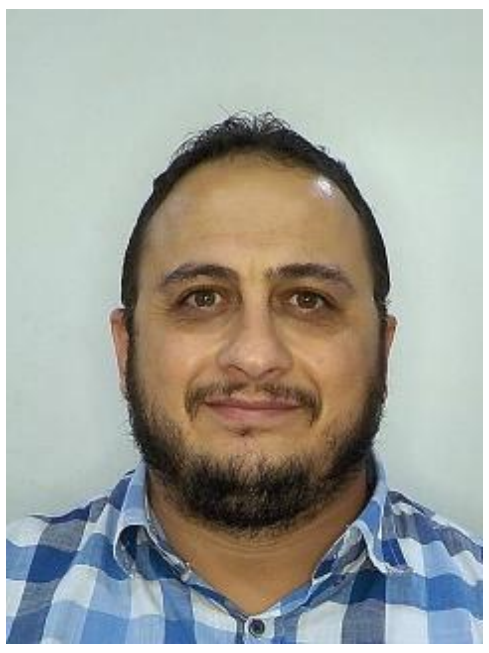

***Mohammad Abufadda*** Holds BSc. in Computer Science, and Currently MSc. Students in Computer Science in Zarqa University, Jordan. Working Currently as programmer in Ministry of Education, Jordan.